\documentstyle[12pt,fleqn]{article}
\newcommand{\bD}{{\bf D}}
\newcommand{\bE}{{\bf E}}
\newcommand{\tD}{\widetilde{\bf D}}
\newcommand{\tE}{\widetilde{\bf E}}
\newcommand{\tph}{\widetilde{\phi}}

\date{}
\title{\bf   
Generation of a dipole moment by external field
in Born-Infeld non-linear electrodynamics}
\author{\\ \\ Dariusz Chru\'sci\'nski\footnotemark\\
        Fakult\"at f\"ur Physik, Universit\"at Freiburg\\
        Hermann-Herder-Str. 3, D-79104 Freiburg, Germany        
  \\  \\  
Jerzy Kijowski\\ Centrum Fizyki Teoretycznej PAN\\
        Aleja Lotnik\'ow 32/46, 02-668 Warsaw, Poland}

\begin{document}

\def\thefootnote{\relax}\footnotetext{$^*$On leave from
 Institute of Physics,
        Nicholas Copernicus University,
        ul. Grudzi\c{a}dzka 5/7, 87-100 Toru\'n, Poland.}

\maketitle
 
\begin{abstract}

The mechanism for the generation of a dipole moment due to an external 
field is presented for the Born-Infeld charged particle. The 
``polarizability coefficient'' is calculated: $\kappa = 1.85407\ r_0^3$, 
where $r_0 := \sqrt{|e| / 4\pi b}\ $ and $b$ is the Born-Infeld
non-linearity constant. Some physical implications are briefly discussed.

\end{abstract}

\vspace{4cm}
\noindent Freiburg THEP-97/31

\newpage

\section{Introduction}

Recently, one of us has proposed a consistent, relativistic theory of
the classical Maxwell field interacting with classical, charged,
point-like particles (cf. \cite{EMP}). For this purpose an ``already
renormalized'' formula for the total four-momentum of a system composed
of both the moving particles and the surrounding electromagnetic field
was used.  It was proved, that the conservation of the total
four-momentum defined by this formula is equivalent to a certain
boundary condition for the behaviour of the Maxwell field in the
vicinity of the particle trajectories. Without this condition, Maxwell
theory with point-like sources is not dynamically closed: initial
conditions for particles and fields do not imply the evolution of the
system because the particle trajectories can be {\em arbitrarily}
chosen and then the initial-value-problem for the field can be solved.
The condition, called {\it fundamental equation}, added to Maxwell
equations, provides the missing dynamical equation: now, particles
trajectories cannot be chosen arbitrarily and initial data uniquely
imply the future and the past evolution of the composed ``particles +
fields'' system.

Physically, the ``already renormalized'' formula for the total
four-mo\-men\-tum was suggested by a suitable approximation procedure
applied to an extended-particle model. In such a model we suppose that
the particle is a stable, soliton-like solution of a hypothetical
fundamental theory of interacting electromagnetic and matter fields.
We assume that this hypothetical theory tends asymptotically to the
linear Maxwell electrodynamics, in the weak field regime (i.e.~for
weak electromagnetic fields and ``almost vanishing'' matter fields).
This means, that ``outside of the particles'' the entire theory reduces
to Maxwell electrodynamics. Starting from this model, a formula was
found, which gives in a good approximation the total four-momentum of
the system composed of both the moving (extended) particles and the
surrounding electromagnetic field.  This formula uses only the
``mechanical'' information about the particle (position, velocity, mass
$m$ and the electric charge $e$) and the free electromagnetic field
outside of the particle. It turns out, that this formula does not
produce any infinities when applied to the case of point particles,
i.e.~it is ``already renormalized''.  We take it as a starting
point for a mathematically self-consistent theory of {\it point-like
particles} interacting with the linear Maxwell field. The ``fundamental
equation'' of the theory is precisely the conservation of the total
four momentum of the system ``particles + fields'', defined by this
formula.

At this point a natural idea arises, to construct a
``second-generation'' theory, which approximates better the real
properties of an extended particle, and takes into account also
possible deformations of its interior, due to strong external field. A
possible description of such a deformation would consist in taking into
account a possible {\it polarization} of the particle,
i.e.~generation of an electric dipole.

The goal of this paper is to illustrate this phenomenon on a simple,
physical model of the ``internal non-linearity'', namely the
Born-Infeld non-linear electrodynamics.

The mathematical mechanism responsible for such a polarizability would
be following.  Given a particular model of the matter fields
interacting with electromagnetism, the ``particle at rest''-solution
corresponds to a minimum of the total field energy. It is, therefore,
described by a solution of a system of elliptic equations
(Euler-Lagrange equations derived from the total Hamiltonian of the
hypothetical fundamental theory of interacting matter fields and
electromagnetic field). This solution corresponds to the vanishing
boundary conditions at infinity.  Physical situation ``particle in a
non-vanishing external field'' corresponds to the solution of the same
elliptic problem with constant, {\em non-vanishing} boundary conditions
${\bf E}(\infty)$ at infinity. Far away from the particle, such a
solution is again a solution of the free Maxwell equations. With
respect to the unperturbed situation, it may contain additional
multipole moments. For weak perturbations, the relation between the
dipole moment ${\bf d}$ created this way and the external field
${\bf E}(\infty)$ is expected to be linear (in the first approximation):
\begin{equation}
{\bf d} = \kappa {\bf E}(\infty)
\end{equation}
and the coefficient $\kappa$ describes the ``deformability'' of the
particle, due to non-linear character of the interaction of the matter
fields (constituents of the particle) with the electromagnetic field.

The coefficient $\kappa$ arises, therefore, similarly as ``reflection''
or ``transmission'' coefficients in the scattering theory. Constant
field and the dipole solution are two independent solutions of the
second order, linear equations describing the free Maxwell field
surrounding the particle. Outside of the particle they may be mixed in
an arbitrary proportion. Such an arbitrary mixture is no longer
possible if it has to match an exact solution of non-linear equations
describing the interior of the particle.

In the present paper we assume that the exterior of the unperturbed
particle is described by the spherically symmetric, static solution of
non-linear Born-Infeld electrodynamics with a $\delta$-like source.
For any value of ${\bf E}(\infty)$ we find explicitly the
two-dimensional family of all the dipole-like perturbations of the
above solution.  They all behave correctly at $r \rightarrow \infty$.
For $r \rightarrow 0$, however, there is one perturbation which remains
regular, and another one which increases faster then the unperturbed
solution. The variation of the total field energy due to the latter
perturbation is infinite, which we consider as an unphysical feature.
We conclude that all the physically admissible perturbations are
proportional to the one which is regular at $0$. At $r \rightarrow
\infty$ this solution behaves as a mixture of the constant field
${\bf E}(\infty)$ and the dipole solution. We calculate the ratio
between these two ingredients and we interpret it as the
polarizability coefficient of the Born-Infeld particle.

\section{Perturbations of the Born-Infeld particle}

The Born-Infeld electrodynamics \cite{BI} (see also \cite{IBB}) is
defined by the following Lagrangian
\begin{eqnarray}    \label{Lag-BI}
{\cal L}_{BI} := b^2\left[1- \sqrt{1-2b^{-2}S - b^{-4}P^2}\right]\ ,
\end{eqnarray}
where the Lorentz invariants $S = \frac 12 ({\bf E}^2-{\bf B}^2)$ and
$P={\bf E}\cdot{\bf B}$ are built of the components ${\bf E}$ and ${\bf
B}$ of the electromagnetic field $f_{\mu\nu}$, defined in a standard
way {\it via} a four-potential vector. The parameter ``$b$'' has a
dimension of a field strength (Born and Infeld called it the {\it
absolute field}, cf. \cite{BI}) and it measures the nonlinearity of the
theory. In the limit $b \rightarrow \infty$ the Lagrangian ${\cal
L}_{BI}$ tends to the standard Maxwell Lagrangian
\begin{eqnarray}     \label{Lag-Max}
{\cal L}_{Maxwell} = S\ .
\end{eqnarray}
Field equations derived from (\ref{Lag-BI}) have the same form as
Maxwell equations derived from (\ref{Lag-Max}), but the relation
between fields ({\bf E},{\bf B}) and ({\bf D},{\bf H}) is now highly
nonlinear:
\begin{eqnarray}
{\bf D} &:=& \frac{\partial {\cal L}_{BI}}{\partial {\bf E}} =
\frac{ {\bf E} + b^{-2}({\bf E}{\bf B}){\bf B} }
{\sqrt{ 1 - b^{-2}({\bf E}^2 - {\bf B}^2) - b^{-4}({\bf E}{\bf B})^2
}}\ , \label{D} \\
{\bf H} &:=& -\frac{\partial {\cal L}_{BI}}{\partial {\bf B}} =
\frac{ {\bf B} - b^{-2}({\bf E}{\bf B}){\bf E} }
{\sqrt{ 1 - b^{-2}({\bf E}^2 - {\bf B}^2) - b^{-4}({\bf E}{\bf B})^2 }}\ .
\label{H}
\end{eqnarray}
The above formulae are responsible for the nonlinear character of the
Born-Infeld theory. In the limit $b\rightarrow \infty$ we obtain linear
Maxwell relations: ${\bf D}={\bf E}$ and ${\bf H}={\bf B}$ (we use the
Heaviside-Lorentz system of units).

Now,  consider a point-like, Born-Infeld  charged particle at rest.
It is described by the static solution of the Maxwell field equations
with $\rho=e\delta({\bf r})$ and ${\bf j}=0$,
where $e$ denotes the particle's electric charge.  Obviously ${\bf
B}={\bf H}=0$ (Born-Infeld electrostatics). Moreover, the spherically
symmetric solution of $\nabla\cdot{\bf D}=e\delta({\bf r})$ is given by
the Coulomb formula
\begin{eqnarray}  \label{D0}
{\bf D}_0 = \frac{e}{4\pi r^3}{\bf r}\ .
\end{eqnarray}
Using (\ref{D}) one easily finds
the corresponding ${\bf E}_0$ field
\begin{eqnarray}    \label{E0}
{\bf E}_0 = \frac{{\bf D}_0}{\sqrt{1+b^{-2}{\bf D}_0^2}} =
\frac{e}{4\pi r}\frac{{\bf r}}{\sqrt{r^4+r_0^4}}\ ,
\end{eqnarray}
where $r_0:= \sqrt{{|e|}/{(4\pi b)}}$ (note, that the field ${\bf
E}_0$, contrary to ${\bf D}_0$, is bounded in the vicinity of the
particle. It implies that the energy of a point charge is already
finite).

Consider a perturbation $\bE = \bE_0 + \tE$ of the static Born-Infeld
solution $\bE_0$,
where $\tE$ is weak perturbation ($|\tE|\ll |\bE_0|$).
The corresponding expansion $\bD :=\bD_0 + \tD + O(\tE^2)$,
where $O(\tE^2)$ denotes terms  vanishing for $|\tE|\rightarrow 0$ like
$\tE^2$ or faster and
\begin{eqnarray}   \label{perturb-D}
\tD = \frac{1}{\sqrt{1-b^{-2}\bE_0^2}} \left( b^{-2}(\bD_0\tE)\bD_0
+ \tE \right)
 = \sqrt{1+b^{-2}\bD_0^2}
 \left( b^{-2}(\bD_0\tE)\bD_0 + \tE \right)
\end{eqnarray}
is the {\it linear perturbation} may be obtained from (\ref{D}). Due to
Maxwell equations we have: $\nabla \times \tE=0$
i.e.~$\tE=-\nabla\tph$.  Therefore, using (\ref{perturb-D}), equation
$\nabla\cdot\tD=0$ implies:
\begin{eqnarray}  \label{tilde-phi}
\Delta\tph + \left( \frac{r_0}{r}\right)^4 \left[
  \frac{\partial^2
\tph}{\partial r^2}
-\frac4r   \frac{\partial\tph}{\partial r} \right]
= 0\ ,
\end{eqnarray}
where $\Delta$ stands for a 3-dimensional Laplace operator in ${\bf
R}^3$ (in the limit $r_0\rightarrow 0$ we obtain simply the Laplace
equation $\Delta\tph=0$ for the electrostatic potential $\tph$). Using
spherical coordinates in ${\bf R}^3$ the Laplace operator $\Delta$ may
be decomposed into radial derivatives and the Laplace-Beltrami operator
${\bf L}^2$ on the unit sphere. Using also the multipole expansion for
the value of $\tph$ on each sphere $\{ r =$const$\}$ we see, that due
to the spherical symmetry of the unperturbed solution $\bD$, different
multipole modes decouple in the above equation. Let us, therefore,
concentrate on the dipole-like deformations, i.e.~on solutions of the
type:
\begin{equation}             \label{tph}
\tph(r,\mbox{angles}) := \Psi(r)\ {\cal E}_kx^k\ ,
\end{equation}
where ${\cal E}_k$ is constant.
Obviously,  ${\bf L}^2\tph=-2\tph$.
 Inserting the above {\it ansatz} to
(\ref{tilde-phi}) we obtain:
\begin{equation}   \label{Psi}
\left(r^2\Psi'' + 4r\Psi'\right) + \left(\frac{r_0}{r}\right)^4
\left(r^2 \Psi'' - 2{r}\Psi' - 4\Psi \right) = 0\ ,
\end{equation}
where $\Psi'$ stands for $\partial\Psi/\partial r$.
Equation (\ref{Psi}) may be explicitly solved (see Section 3).

For $r \gg r_0$, equation (\ref{tilde-phi}) reduces to the standard
Laplace equation with 2 independent dipole-like solutions: the constant
one and the external dipole solution. Correspondingly, there is a
solution $\Psi_1$ of (\ref{Psi}) which is constant at infinity and a
solution $\Psi_2$ which behaves like $r^{-3}$ at infinity. They can be
chosen as a basis in the space of solutions and any other solution is
of the form
\begin{equation}  \label{solution2}
\Psi(r) =
A\Psi_1(r) + B\Psi_2(r) \ .
\end{equation}
On the other hand, a basis may be chosen corresponding to the behaviour
of $\Psi$ at $r \rightarrow 0$. There is a solution $\Psi_3$, which
is bounded (it behaves like $r^4$) and another one, say $\Psi_4$, which
behaves like $r^{-1}$.  Correspondingly, any solution may be also
represented as a combination
\begin{equation}  \label{solution1}
\Psi(r) =
C\Psi_3(r) + D\Psi_4(r) \ .
\end{equation}
Physical assumptions impose condition $D = 0$, because otherwise the
``perturbation'' $\tE$ would become infinite. Such a behaviour
contradicts the universal condition $|{\bf E}| < b$, implied by
equation (\ref{D}) and proves that the corresponding solution is merely
an artefact of the linearization. Moreover, it leads to an infinite
variation of the total field energy (see Section~3).  This reduces the
space of possible solutions to one dimesion and imposes the linear
relation between the coefficients $A$ and $B$ in the representation
(\ref{solution2}), because we have
\begin{equation}  \label{solution}
A\Psi_1(r) + B\Psi_2(r)  =  C\Psi_3(r) \ .
\end{equation}
Hence, we may define the
``deformability coefficient'':
\begin{equation}
\kappa := \frac BA \ ,
\end{equation}
equal to the ratio between the dipole moment ${\bf d}$ generated by
the external field $\tE (\infty)$ and the field itself.
Using condition (\ref{solution}) and its derivative at some fixed point
$r_0$, we easily eliminate the arbitrary variable $C$ and obtain the
following result:
\begin{equation}    \label{kappa}
\kappa  =  \frac{\Psi'_1(r_0)\Psi_3(r_0) -
\Psi_1(r_0)\Psi'_3(r_0)}{\Psi'_2(r_0)\Psi_3(r_0) -
\Psi_2(r_0)\Psi'_3(r_0)} \ .
\end{equation}
In the next Section we show that
\begin{equation}   \label{kappa-num}
\kappa = - 1.85407\ r_0^3\ .
\end{equation}
Obviously, in the limit of Maxwell theory ($r_0\rightarrow 0$), the
external electric field does not generate any particle's dipole moment.
Therefore, this mechanism comes entirely from the non-linearity of the
Born-Infeld theory.

\section{Exact solutions of the linearized Born-Infeld electrostatics}

To find solution $\Psi_1$ let us introduce the
variable
$z:=-\left(\frac{r_0}{r}\right)^4$.
 Then equation (\ref{Psi}) reduces to the hypergeometric equation (see
e.g. \cite{Watson}):
\begin{equation}                \label{Psi1}
z(1-z) \frac{d^2\Psi_1}{dz} + \left(\frac 14 - \frac 74 z\right)
\frac{d\Psi_1}{dz} + \frac 14 \Psi_1 = 0\ .
\end{equation}
Its regular solution is given by
\begin{equation}
\Psi_1(z) = F\left(-\frac 14,1,\frac 14,z\right)\ ,
\end{equation}
where $F(a,b,c,z)$ denotes a hypergeometric function (cf.
\cite{Watson}).

To find $\Psi_2$ let us define a function $\Phi_2:=r^{3}\Psi_2$.  Note,
that $\Phi_2$ is constant at infinity. The equation (\ref{Psi})
rewritten in terms of $\Phi_2$ also reduces to the hypergeometric
equation in the variable ``$z$''. Its regular solution reads:
\begin{equation}
\Phi_2(z) = F\left(\frac 74,\frac 12, \frac 74,z\right)\ .
\end{equation}

Another way to parameterize the space of solution of (\ref{Psi}) is to
take as a basis 2 independent solutions $\Psi_3$ and $\Psi_4$ which
behave for $r\rightarrow 0$ like $r^{4}$ and $r^{-1}$ correspondingly.
Let us first look for $\Psi_4$ and define $\Phi_4 := r\Psi_4$. From
(\ref{Psi}) one gets the hypergeometric equation in the variable
$w:=1/z$ with the solution
\begin{equation}
\Phi_4(w)= F\left(\frac 12,-\frac 14,-\frac 14,w\right)\ .
\end{equation}
We shall prove that the solution $\Psi_4(r)=r^{-1}\Phi_4(-(r/r_0)^4)$ is
unphysical. For this purpose let us analyze the behaviour of
the electric field
$\tE$
``produced'' by $\Psi_4$  in the vicinity of the Born-Infeld
particle, i.e. for $r\rightarrow
0$. From (\ref{tph}) we get
\begin{eqnarray}    \label{singular-E}
\widetilde{E}_k =
- \partial_k\tph \ \sim \ \frac 1r  \left( \delta^l_{\ k} -
\frac{x^lx_k}{r^2} \right) {\cal E}_l   + \frac{r^3}{2r_0^4}
\left(\delta^l_{\ k} +
3\frac{x^lx_k}{r^2}\right){\cal E}_l + O(r^7)\  ,
\end{eqnarray}
The first term in (\ref{singular-E}), $\tE$  exceeds the
unperturbed field $\bE_0$ itself.

Moreover, the ``perturbation'' $\tE$ leads to infinite variation of the
total field energy. The ``electrostatic'' energy ${\cal H}$
corresponding to electric induction {\bf D} is given by (cf. \cite{BI},
\cite{IBB}):
\begin{eqnarray}
{\cal H} = \int b^2\left( \sqrt{1+b^{-2}{\bf D}^2} - 1\right)d^3x\ .
\end{eqnarray}
Therefore, its variation reads
\begin{eqnarray}
\delta {\cal H}|_{\bD_0} = \int  \frac{D^k_0\delta
D_k}{\sqrt{1+b^{-2}{\bf D}^2_0}}
 \,d^3x\
= \int   E^k_0\delta D_k\, d^3x =
\int   E^k_0   \widetilde{D}_k\,  d^3x\ .
\end{eqnarray}
Now, let us analyze the behaviour of
$\bE_0\tD$ for $r\rightarrow 0$.
It follows from (\ref{perturb-D}) that this expression contains a
highly singular term which behaves like $r^{-6}\times \mbox{radial\
component\ of}\ \tE$. Using expansion (\ref{singular-E}) we see that
the first term is purely tangential, i.e. it is orthogonal to ${\bf
D}_0$.  However, the second term does contain a radial part $\sim
r^3{\cal E}^r$ and this way $\bE_0\tD$ produces a nonintegrable
singularity $r^{-3}{\cal E}^r$.  We conclude that the
solution $\Psi_4$ is unphysical.

Finally, to find $\Psi_3$ we define a function $\Phi_3:= r^{-4}\Psi_3$
which fulfils once again the hypergeometric equation in the variable
$w$. Its solution is given by
\begin{equation}
\Phi_3(w)= F\left(\frac 74,1,\frac 94,w\right)\ .
\end{equation}

To calculate the ``deformability coefficient'' $\kappa$ from a formula
(\ref{kappa}) one needs also the derivatives of $\Psi_1$, $\Psi_2$ and
$\Psi_3$. However, it is easy to compute them using the following well
known formula for a hypergeometric function (cf. \cite{Watson}):
\begin{equation}
\frac{d}{dz}F(a,b,c,z) = \frac{ab}{c}F(a+1,b+1,c+1,z)\ .
\end{equation}
Therefore, taking into account numerical values of corresponding
hypergeometric functions
one finally gets  (\ref{kappa-num}).

One might describe, this way, the polarizability coefficient of the
proton. According to \cite{Bar}, we have $\kappa = (12.1 \pm 0.9)
\times 10^{-4}\ \mbox{fm}^3 $. To fit this value one has to take
$r_0 \approx 0.09\ \mbox{fm}$. The total mass of the corresponding Born-Infeld
unperturbed field acompanying such a particle is about 32 electron
masses. We see, that the main part of the proton total mass cannot be
of electromagnetic nature and has to be concentrated in the material
core of the particle.

\section*{Acknowledgement} D.C. thanks Alexander von 
Humboldt Stiftung for the financial support.

\end{document}